# Boulders on Mercury

*Mikhail A. Kreslavsky[1], Anastasia Yu. Zharkova[2,3], James W. Head[4], Maria I. Gritsevich[5,6,7]*

[1]Earth and Planetary Sciences, University of California, Santa Cruz, CA, 95064 USA, mkreslav@ucsc.edu

[2]MExLab, Moscow State University of Geodesy and Cartography (MIIGAiK), Moscow, Russia.

[3]Sternberg Astronomical Institute, Moscow State University, Moscow, Russia

[4]Earth, Environmental and Planetary Sciences, Brown University, Providence, RI, 02912, USA

[5]Finnish Geospatial Research Institute (FGI), Masala, Finland

[6]Department of Physics, University of Helsinki, Finland

[7]Institute of Physics and Technology, Ural Federal University, Ekaterinburg, 620002, Russia.

**Abstract**

Boulders on the surfaces of planets, satellites and small bodies, as well as their geological associations, provide important information about surface processes. We analyzed all available images of the surface of Mercury that have sufficient resolution and quality to detect boulders, and we mapped all the boulders observed. The lower size limit of detectable boulders was ~5 m. All boulders found on Mercury are associated with fresh impact craters hundreds of meters in diameter or larger. We compared boulder population on Mercury with population of boulders of the same size on the Moon, and found that boulders on Mercury are ~30 times less abundant than in the lunar highlands. This exact quantitative estimate is inherently inaccurate due to the limitation in the source data; however, the significant relative rarity of boulders on Mercury can be firmly and reliably established. We discuss possible causes of the observed difference. Higher thermal stresses and more rapid material fatigue due to diurnal temperature cycling on Mercury may cause rapid disintegration of the upper decimeters of the boulder surface and thus contribute to more rapid boulder obliteration; however, these factors alone cannot account for the observed difference. A proposed thicker regolith on Mercury is likely to significantly reduce boulder production rate. A higher micrometeoritic flux on Mercury is likely to result in micrometeoritic abrasion being a dominant contributor to boulder degradation; this high abrasion rate likely shortens the boulder lifetime. A combination of these factors appears to be able to account for the relative rarity of boulders on Mercury.

*Highlights:*
- We searched all available very-high-resolution images of Mercury for boulders
- Boulders on Mercury are extremely rare in comparison to the Moon
- This is caused by a thicker regolith and disproportionally higher micrometeorite flux
- More intensive thermal cracking and thermal fatigue may also play some role

# 1. Introduction

Boulders, meters-size blocks of solid rock, are seen in great numbers in high-resolution images of the surfaces of the Moon (Figure 1), Mars, small bodies, as well as the Earth. Boulders are observed on all solar system bodies, where images of sufficient resolution are available. The presence of boulders and their geological associations provide information about surface properties and modification processes on these bodies. For example, Bart & Melosh (2010), Krishna & Kumar (2016), Pajola et al. (2018) have considered distribution of boulders ejected by impacts on the Moon with implications for the impact mechanics. Bart & Melosh (2007) have shown that size of boulders excavated by kilometer-size craters on the Moon might be used to distinguish primary and secondary impacts. Traces of recent boulder falls have been used as indication of recent seismic activity on the Moon (e.g., Kumar et al. 2016, 2019a) and Mars (Kumar et al., 2019b). On Mars, studies of boulders have shed light on periglacial processes (e.g., Orloff et al., 2011, 2013; Levy et al., 2018). In a number of works boulders have been studies as landing hazards. Of course, this brief overview of planetary boulder studies is far from complete. As far as we know, boulders on Mercury have not been studied yet.

On Mercury images of sufficient resolution to see boulders are rare. We searched all high-resolution images for boulders and report our results here. We found that boulders on Mercury are very rare in comparison to the Moon. We discuss several possible causes of this difference between the two bodies and present relevant order-of-magnitude estimates pertaining to each factor. A preliminary version of this work has been reported by Zharkova et al. (2019).

The next section describes images of Mercury that we used as well as our approach of selecting and simulating lunar image samples of the same quality for quantitative comparison. In Section 3 we present our observational results. In Section 4 we discuss possible causes of the observed difference between the Moon and Mercury. Finally, in the concluding section we summarize our findings and briefly discuss prospects for future studies.

# 2. Data and Methods

## 2.1. Boulders on Mercury

To study boulders on Mercury we used images obtained by the Narrow Angle Camera (NAC) of the Mercury Dual Imaging System (MDIS) instrument (Hawkins et al., 2007; Denevi et al., 2018) onboard the MErcury Surface, Space Environment, Geochemistry and Ranging (MESSENGER) during its orbital mission to Mercury. We selected images of the highest resolution and the most detailed sampling (better than 2.5 m/pix, informally called "high-resolution images" hereafter) acquired from February to April of 2015, during the late phase of the mission, when MESSENGER was on a low-periapsis orbit, and the MDIS NAC camera was operating under imaging conditions well beyond the limits foreseen by its design. Individual high-resolution images are small (0.25 Mpix), have a considerable amount of smear (because of the short range to the surface), low signal-to-noise ratio (because of the short exposures needed to keep image smear reasonable), and do not overlap: the distance between consecutive images within the same orbit is much greater than the image size (~0.5 – 1 km). These images cannot be used to produce mosaics, unlike the regular (10 – 20 m/pix) operational mode of the NAC. Moreover, they cannot be placed in the context of lower-resolution images because of the large difference in resolution and small image size. In a way, the high-resolution images are random samples of Mercury surface, informally referred to as "postage stamps".

We systematically screened all ~3000 nadir-looking high-resolution MDIS NAC images with smear less than 10 pixels, and sunlight incidence angle less than 70° (to avoid images with large areas of shadows). All these images are located in a region between 40 – 70°N and 210 – 320°E. This region is dominated by the intercrater plains (Denevi et al., 2013; Whitten et al., 2014) (Figure 2). One of us (AYuZh) screened the whole image set (thus ensuring the same boulder identification criteria) and identified all discernable boulders. Only 14 individual images with boulders were found. Another one of us (MAK) independently screened about a third of the image set and obtained identical results with no discrepancies (which confirms the objectiveness of the survey). Examples of images in which boulders were found are shown in Figure 3; locations of images with identified boulders are shown in Figure 2 and listed in Table 2, and a summary is presented in Table 1.

The boulders often are at the resolution limit. Because of their small size, the smear and low signal-to-noise ratio, it is difficult to measure boulder size accurately. We estimate that our boulder detection limit is ~4 – 5 m; this value varies slightly between images due to differences in pixel size and noise level. The largest boulder in Fig. 1 would be discernable in the surveyed images. With the available images, it is impossible to distinguish between intact boulders and debris piles of disintegrated or partly disintegrated boulders, therefore such (partly) disintegrated boulders are included in our boulder count.

**2.2. Boulders on the Moon**

To compare boulders on Mercury with those in the lunar highlands, we used images obtained by Lunar Reconnaissance Orbiter Camera (LROC) Narrow Angle Camera(NAC) (Robinson et al., 2010). Those images are large, have a high signal-to-noise ratio, and are numerous. Due to their high quality, much smaller boulders are seen in them than in the MESSENGER images. For quantitative comparison we simulated the MDIS NAC high-resolution images set. We created a random sample of lunar images with the same distribution of image resolution, image size, image quality and illumination / observation conditions as the source MDIS image set. To do this, for each high-resolution MDIS image surveyed we ran an automated procedure that performed the following steps. It found all LROC images of the same pixel size (within ± 10%) and the same solar incidence angle (within ± 0.5°). From all the highland area covered with such found images it selected a random point. It extracted a portion of the LROC image with this point and projected it to local orthographic projection (which provides a good proxy for MDIS imaging geometry). Then it degraded image quality: it applied isotropic Gaussian blur with 1.5 pixel width, and smeared the image for known smear length in the direction making the same angle with the illumination direction as in the source MDIS image. The latter is important, because the effect of the smear on boulder visibility strongly depends on this angle; for example, if smear is parallel to the illumination direction, a bright boulder and its dark shadow effectively cancel each other and the boulder disappears; fortunately, these directions were not parallel on the source MDIS images. Both filtering procedures (blur and smear) were implemented as linear filters in the spatial domain. Then our automated procedure added white uniform noise to reduce the signal-to-noise ratio; the noise amplitude was the same for all images; it was chosen through visual comparison of several trial MDIS and processed LROC images. Finally, a 0.25 Mpix sample was extracted from the processed LROC image.

We limited our survey of the random lunar samples to ~380 randomly chosen samples, because further increase of sample number would not increase the accuracy of the results, which are limited by the Mercury data (see Section 3.1 below). All actual boulder identification on the lunar sample images was done by AYuZh, which ensures identical boulder identification criteria on both the Moon and Mercury. The summary of the random sample survey is shown in Table 1.

## 3. Results

### 3.1. Rarity of boulders on Mercury

The random nature of surface sampling with the high-resolution MDIS images and the lunar sample images enables a quantitative comparison of boulder abundance on the Moon and Mercury. Boulders often occur in clusters and large fields both on Mercury and the Moon. On the other hand, the majority of high-resolution images on Mercury and image samples on the Moon do not contain boulders. Due to these two facts, the mean boulder density (the number of documented boulders divided by the surveyed area) is not a good statistical measure of a typical boulder abundance: it is not statistically stable because its value would be dominated by infrequent, occasionally encountered dense boulder clusters. The percentage of random image samples with boulders among the entire image set is a better statistical characteristic of the boulder abundance, it is more statistically stable and, therefore, more objective. Table 1 shows that according to this measure, boulders on Mercury are extremely rare in comparison to the lunar highlands, about 30 times less abundant (~0.5% vs. ~15%). Boulders are more frequent in lunar maria than in highlands, therefore the difference between Mercury and lunar maria is even greater.

Unfortunately, the quantitative Mercury/Moon boulder abundance ratio obtained is not accurate. Its statistical uncertainty is dominated by a small number of images with boulders on Mercury and cannot be improved, because no additional suitable images exist. The formal Poissonian 90% confidence interval for the percentage is 0.32% – 0.63%, almost a factor of two difference. The actual accuracy is even worse, because sampling is not perfectly random, and the abundance measure is not perfectly statistically stable: the number of images with boulders depends on how a single young large crater is represented in the image set, see Section 3.2 below. Nevertheless, we can conclude that our survey demonstrates with great certainty that boulders on Mercury are very rare in comparison to the Moon.

### 3.2. Geologic associations of boulder occurrence

Five (5) of the 14 high-resolution Mercury images with boulders are consecutive images from a single MESSENGER orbit (Figure 2). These images are located in and near a large (~35 km in diameter) unnamed impact crater centered at 64.5°N 104.6°W (Figure 4). Four (4) of them each contain many boulders (Figure 3a). MDIS WAC images (Figure 4a) show that the large crater appears bright, fresh and sharp, and that it is associated with several chains of small craters with sharp, crisp morphology. An enhanced color mosaic of WAC images (Denevi et al., 2018) (available at https://messenger.jhuapl.edu/Explore/Science-Images-Database/gallery-image-1226.html) shows blue (high violet-to-IR spectral ratio) rays associated with the crater. Regular MDIS NAC images (Figure 4b) and high-resolution NAC images show sharply defined impact melt pools on the outer crater walls and crisp cracks in the impact melt on the crater floor. All these features indicate that the crater is very young, among the youngest craters of its size on the planet, and the youngest large crater in the area with high-resolution images (Zharkova et al.,

2020). Banks et al. (2017) have listed this crater as Kuiperian, which is consistent with our analysis. Analogously, on the Moon abundant boulders and large blocks are always seen in the young (Copernican age) large craters (e.g., Hörz et al, 2020, and references therein).

The remaining nine (9) of the 14 Mercury images with boulders are scattered across the sampled area. Three (3) of them are in the smooth plains (Denevi et al., 2013) and the other six (6) are in the intercrater plains. The proportion of images is shifted toward smooth plains in comparison to the sampling proportion; however, the statistics are insufficient to claim a significant preference for smooth plains. Seven (7) of these Mercury images contain only one boulder each. All boulders in these nine images are apparently associated with relatively fresh impact craters hundreds of meters in diameter (Figure 3b). The ages of these small craters are difficult to assess; however, the craters are relatively deep in comparison to heavily degraded craters of similar size, which indicates their relative youthfulness. The association of boulders with small fresh craters is also typical to the Moon (e.g., Hörz et al, 2020, and references therein). Obviously, such boulders were ejected by the crater-forming impacts (e.g., Bart and Melosh, 2010; Krishna and Kumar, 2016).

On the Moon, boulders also often occur in other geological settings that are distinctly different from those of young large or small impact craters. These include hilltops and other convex slope breaks occurring in association with a variety of features of Imbrian and Eratosthenian ages. In these geological environments, diffusive regolith creep (Soderblom, 1970; Fassett and Thomson, 2014) effectively removes regolith and continues to expose megaregolith blocks and bedrock outcrops. For example, the boulder in Fig. 1 rolled down to its present location from such a geological setting in the upper, convex part of the south-eastern slope of the North Massif at the Apollo 17 landing site. Regolith thinning, bedrock exposure, and abundant small boulders have all been field-documented for the upper edge of sinuous rille Rima Hadley at Apollo-15 landing site (Swann et al., 1972), as well as for the upper edge of a linear rille informally known as Fossa Recta in crater Le Monnier with Lunokhod-2 rover data (Basilevsky et al., 1977). In our random lunar image samples, we see a few unambiguous examples of boulder fields in such settings (although often it is difficult to classify boulder occurrence settings because samples are too small to determine the context).

We have not found any boulder in such geologic settings on Mercury: all boulders have apparent association with young impact craters. This could occur just by chance because only as few as nine scattered images have boulders. This also might be related to the fact that the intercrater plains on Mercury are significantly smoother than the lunar highlands (Kreslavsky et al., 2014; Whitten et al., 2014); in this case, suitable hilltops and slope breaks are predicted to be less abundant and might not be sampled in our limited image set. It is also possible that a thicker regolith and a higher regolith production rate on Mercury (see discussion and references in sections 4.1, 4.2 below) effectively reduce bedrock exposure in such settings.

## 4. Discussion: causes of the rarity of boulders on Mercury

On the Moon, the boulder population is dynamic: boulders are constantly being produced by meteoritic impacts and excavated from the hill tops and other convex slope breaks by regolith creep and are constantly being obliterated. On an observational basis, Basilevsky et al. (2013, 2015) estimated the characteristic lifetime of lunar boulders to be on the order of 100 Ma. The possible mechanisms of boulder obliteration (described in Horz et al. 2020) are (1) meteoritic impacts, (2) micrometeoritic abrasion and (3) "thermal fatigue", material fatigue due to incremental crack propagation induced by diurnal thermal stress cycling. All three mechanisms undoubtedly contribute to boulder obliteration, but their contributions are not likely to be similar in magnitude. On the basis of model calculations (Hörz et al., 1974, 1975), Hörz et al. (2020) concluded that micrometeoritic abrasion is an order of magnitude less effective than meteoritic impacts, thus making only a minor contribution. Hörz et al. (2020) presented several lines of observational evidence for meteoritic impacts rather than thermal fatigue being the dominant boulder obliteration factor on the Moon. Hörz et al. (1975) and Basilevsky et al. (2013) successfully reproduced the observed boulder lifetime with a model of boulder disintegration by impacts.

It is very reasonable to assume that the boulder population on Mercury is also dynamic. In the dynamic equilibrium, the boulder abundance is equal to the boulder production rate multiplied by boulder lifetime. Therefore, the extreme rarity of boulders on Mercury in comparison to the Moon means that that either boulder production on Mercury is much slower, or boulder lifetime is much shorter, or both. If meteoritic impacts are the main cause of both boulder production and obliteration, the impact flux itself would have little effect on the product of production rate and lifetime and, therefore, on boulder abundance. To explain the observations, we need some major factors that would have disproportional (with respect to the impact rate) effect on boulder production and lifetime. Systematic differences in surface composition and effect of impact velocity on target material fragmentation might affect boulder production and lifetime, but these factors seem to be minor and unable to explain the huge difference between the Moon and Mercury. Below we consider three potential factors that could significantly disproportionally decrease boulder production rate and shorten boulder lifetime on Mercury.

**4.1. The effect of regolith thickness on boulder production rate**

Several lines of indirect evidence suggest that the regolith on Mercury is thicker than on the Moon. Kreslavsky et al. (2014) and Kreslavsky and Head (2015) analyzed the scale dependence of topographic roughness on the Moon and Mercury and concluded that the characteristic spatial scale of terrain smoothing due to regolith gardening on Mercury is a factor of three longer. They interpreted this as a result of a proportionally thicker regolith. Kreslavsky & Head (2015), Zharkova et al. (2015) reported evidence of very thick regolith on Mercury on the basis of morphology of small impact craters. A thicker regolith is a predictable consequence of the higher micrometeoritic flux on Mercury (Cintala, 1992; Borin et al., 2009; see discussion in the next subsection); it is also consistent with the observed higher degradation rate of kilometers-size craters on Mercury (Fassett et al., 2017).

The smallest impact craters are formed entirely in the regolith; they excavate only fragmented material and cannot produce boulders. Formation of boulders requires excavation from the bedrock or megaregolith substrate (e.g., Head and Wilson, 2020). A thicker regolith means that a proportionally larger onset crater size is needed for boulder excavation. This is consistent with the fact that all observed scattered boulders on Mercury are associated with craters that are hundreds of meters in diameter, while in the lunar maria, craters several tens of meters in diameter can produce small boulders. The crater production function (cumulative size-frequency distribution of newly forming craters) for small craters is steep; it can be approximated by a power law with an exponent of ~ -3. A factor of three thicker regolith would mean that boulder-producing impacts would be $\sim 3^3 = 27$ times less frequent.

Taken at its face value, this number appears to coincide with the observed boulder abundance difference; however, taking it at its face value would be incorrect. First, in our measurements of rock abundance, the contribution from the largest craters is greater than the contribution from other craters, because a single large crater can contribute to several randomly sampled images. Second, the smallest boulder-producing impacts on the Moon produce only small boulders, smaller than our detection limit.

In summary, a thicker regolith on Mercury can cause, and indeed is likely to cause, a disproportionally lower boulder production rate, and this decrease is very likely to be significant. However, this factor is unlikely to account for the entire observed rarity of boulders on Mercury in comparison to the Moon.

**4.2. The effect of micrometeoritic impacts on boulder lifetime**

The flux of larger and smaller projectiles producing and destroying boulders is almost proportional. However, this is not the case for micrometeorites responsible for boulder abrasion. The dynamics of projectiles from ~ 1 cm to ~ 5 μm in size is different: the Poynting–Robertson drag plays a significant role in their orbital evolution, causing them to spiral toward the Sun (e.g., Grün, 2007). This process disproportionally increases the number of micrometeoritic impacts closer to the Sun.

Cintala (1992) estimated the magnitude of this effect using analytical techniques. He found that the dust particle impact rate at Mercury is 5.5 times greater, and the mean impact velocity is 1.6 times higher, than at the Moon. Assuming that the abrasion efficiency is proportional to the impact energy, these two results predict that there is an ~14 times enhancement in the abrasion rate on Mercury, relative to the Moon. This means, that in contrast to the Moon, abrasion by micrometeoritic impacts on Mercury plays a significant role in boulder degradation.

Borin et al. (2009, 2017) have argued that the analytical calculations by Cintala (1992) do not take into consideration a number of non-negligible dynamic effects. They performed Monte-Carlo modeling of the orbital evolution of a limited set of small particles and conclude that the Mercury/Moon dust impact flux ratio could be as high as ~54 (Borin et al., 2017), and impact velocities could be 1.1 – 1.4 times higher than on the Moon. These new numbers mean that the micrometeorite abrasion rate on Mercury is two orders of magnitude higher, which in turn means that abrasion not only dominates over larger impacts, but also significantly shortens boulder lifetime on the surface of Mercury.

In summary, micrometeoritic abrasion plays a disproportionally high role on Mercury. It may be a dominant factor in boulder obliteration, and in combination with the effect of a thicker regolith on Mercury, may be able to account for the observed qualitative difference in boulder populations in comparison to the Moon.

**4.3. The role of thermal stresses**

Observed anisotropy of crack orientation in rocks in arid areas on the Earth (McFadden et al., 2005) as well as on Mars (Eppes et al., 2016) has been considered as indirect evidence for a role of thermal fatigue in rock fracturing. The potential role of thermal stresses in boulder disintegration on the Moon and other atmosphereless bodies has been discussed theoretically in great detail by Molaro and Byrne (2012), Molaro et al. (2015, 2017), Delbo et al. (2014). Here we present only a brief theoretical summary.

The effect of thermal stresses is somewhat counterintuitive. From everyday life experience we would expect thermal cracking to be caused by high temperature gradients, such as when boiling water is rapidly poured into a glass. In the natural planetary environment, however, where heating occurs by solar illumination and cooling occurs by thermal radiation, such high local temperature gradients never occur (e.g., Molaro and Byrne, 2012). Theoretical analysis shows that the peak thermal stress is proportional to the diurnal temperature amplitude and does not depend on temperature gradient.

The simplest model of this phenomenon is a homogeneous elastic monolith with a flat surface ($x - y$ plane) and temperature $T = T_0 + \Delta T(z)$ changing as a function of depth $z$ and time of the day, but uniform across the surface. Here $T_0$ is the long-term-average temperature, at which there are no thermal stresses. From the equations of linear elasticity theory (e.g., Landau and Lifshits, 1970; Parkus, 1976), it is easy to derive that such temperature distribution induces stresses described by a stress tensor $\sigma_{ik}$ ($i, k = x, y, z$) with only two non-zero components:

$$\sigma_{xx} = \sigma_{yy} = - E\, \alpha\, \Delta T / (1-\nu), \qquad (1)$$

where $E$ is the Young's modulus ($E \sim 60$ GPa for basalts; e.g., Schultz, 1993), $\nu$ is the Poisson ratio ($\nu \sim 0.25$), and $\alpha$ is the linear thermal expansion coefficient ($\alpha \sim 0.5 \times 10^{-5}$ K$^{-1}$). The minus sign denotes that elevated temperature ($\Delta T > 0$) causes compressional stresses. If a boulder is large in comparison to the diurnal thermal skin depth, this idealized model is applicable to its surface, and Eq. (1) can be used for a crude estimation of a typical thermal stress. If a boulder is small in comparison to the thermal skin, temperature differences inside it are minor, and thermal stresses are negligible. In transitional cases of boulders comparable to the thermal skin depth, stresses are lower than predicted by Eq. (1), but still comparable to them, while stress tensor orientation can be significantly different. Molaro et al. (2017) presented a realistic model of a diurnal cycle of temperature and stresses in boulders comparable to the thermal skin depth.

The thermal skin depth $L$ is

$$L \sim (D_T\, P/2\pi)^{1/2}, \qquad (2)$$

where $D_T \sim 10^{-6}$ m$^2$s$^{-1}$ is thermal diffusivity, and $P$ is the solar day period. For the Moon $L \sim 0.6$ m, and for Mercury, $L \sim 1.5$ m. Therefore, boulders considered in our work are large in comparison to $L$ (the day heat and night cold do not reach their deep interior), and Eq. (1) gives a valid order of magnitude estimate for the diurnal thermal stresses near their surface.

Delbo et al. (2014), Molaro et al. (2015, 2020) noted another mechanism generating thermal stresses. The difference $\Delta\alpha$ between thermal expansion coefficients of rock-forming mineral grains leads to thermal stress $\sigma_{ik}$ at the grain scale. Such a thermal stress tensor is scaled as:

$$\sigma_{ik} = C_{ik}\, E\, \Delta\alpha\, \Delta T, \qquad (3)$$

where spatially varying dimensionless coefficients $C_{ik}$ depend on $\nu$, relative differences in the elastic moduli ($\Delta E/E$), and the grain geometry. For example, for an idealized case of an isolated spherical inclusion in a homogeneous matrix with the same $E$ and $\nu$, the highest stress occurs in the matrix at the inclusion boundary, where the principal stresses (radial stress $\sigma_{rr}$ and hoop stress $\sigma_{\theta\theta} = \sigma_{\phi\phi}$, in the spherical coordinates $r$, $\theta$, $\phi$) are given by Eq. (3) with

$$C_{rr} = -2/3\,(1-\nu),$$
$$C_{\theta\theta} = C_{\phi\phi} = 1/3\,(1-\nu). \tag{4}$$

A positive $\Delta\alpha\,\Delta T$ causes compressional radial stress and tensional hoop stress. For inclusions of a realistic shape, significant stress concentration would occur at sharp inclusion edges and tips, and the absolute values of coefficients $C_{ik}$ may locally be much higher than given by Eq. (4). Values of $\Delta\alpha$ are not small, they are on the same order of magnitude as $\alpha$ itself; for example, Molaro et al. (2015) used $\Delta\alpha = -0.4\times10^{-5}$ $K^{-1}$ for plagioclase inclusions in a pyroxene matrix. Thus, grain-level thermal stresses are of the same order of magnitude as bulk thermal skin stress, and can exceed it locally. Both bulk and grain stresses are proportional to $\Delta T$.

To obtain quantitative stress estimates with Eq. (1), we need to know $\Delta T$. The observed surface temperature pertains to regolith; instead, we need to use modeled temperature for solid rocks. Following Bandfield et al. (2011) we adopt a rock thermal conductivity of 1.49 W m$^{-1}$K$^{-1}$, a density of 2940 kg m$^{-3}$, and a specific heat of 555 J kg$^{-1}$K$^{-1}$, which yields a thermal inertia of 1560 W m$^{-2}$K$^{-1}$s$^{1/2}$. For our order of magnitude estimates, we ignore the temperature dependence of the thermophysical properties and the effects of Mercury eccentricity. These simplifications introduce less inaccuracy than the natural effect of the boulder facet orientation variability. We model the diurnal temperature cycle for a horizontal monolith surface at the equator using KRC code (Kieffer, 2013) (the code is available at http://krc.mars.asu.edu/). Since tensile strength of rocks is much lower than compressional strength, we are interested in the night-time tensional stress, therefore for $\Delta T$ we take the difference between the day-average and day-minimum surface temperatures, which gives $\Delta T \approx 70$ K and $\Delta T \approx 180$ K for the Moon and Mercury, respectively. Thus, thermal stresses on Mercury are a factor of ~2.5 stronger than on the Moon. This difference potentially might trigger much more effective fracturing of boulders due to thermal stresses.

The absolute stress values calculated with Eq. (1) are ~30 MPa and ~ 75 MPa for the Moon and Mercury, respectively. Localized stresses at grain boundaries according to Eq. (3) can be several times higher. In their modeling, Molaro et al. (2015) obtained localized stresses well above 100 MPa for lunar conditions, therefore it is logical to anticipate stresses above 300 MPa on Mercury. These stress values are to be compared with the tensile strength of presumed boulder material, but this is not straightforward, because "strength" is a poorly defined quantity. Small intact basalt samples in the laboratory have a tensile strength of several hundreds of MPa, while for large rock masses, Schultz (1993) recommended a tensile strength value of ~14 MPa as more appropriate. Lakshmi and Kumar (2020) measured tensile strength of a number of basaltic boulders ejected by terrestrial Lonar impact and obtained values of ~10 – 20 MPa for massive boulders and as 3 – 5 MPa for vesicular boulders. Comparison among these numbers indicates that thermal stresses are likely to cause both cracking of boulders at the meter scale, and formation of grain-scale microcracks.

These results, however, do not mean that the boulders would be immediately disintegrated by thermal stresses as soon as they are exposed to the surface environment of Mercury. Small terrestrial rocks placed in a campfire experience $\Delta T$ well above 180 K; they develop microcracks, some rocks in the campfire indeed fall apart; often, however, they remain intact and preserve their strength. Thus, it is not likely that the thermal stresses themselves cause immediate disintegration of the majority of large boulders on Mercury.

Delbo et al. (2014), Molaro et al. (2020) have considered the potential role of thermal fatigue and argued that this mechanism dominates disintegration of decimeter-size rocks on asteroids. Scaling of the Delbo et al. (2014) modeling results for asteroids at 1 AU to the Moon leads to much shorter boulder lifetimes than are observed and thus is inconsistent with observations (see discussion by Hörz et al., 2020). Delbo et al (2014) erroneously used the entire diurnal surface temperature amplitude for $\Delta T$ in their calculations; the correct approach would be to use a factor of two lower difference between the long-term average and one of the extremes. Correcting this would increase the modeled boulder life span by an order of magnitude; while an important factor, this would still be insufficient to explain the discrepancy with observations.

Nevertheless, in the framework of the semi-theoretical treatment of thermal fatigue applied by Delbo et al. (2014), rock lifetime is proportional to $P \Delta T^{-N_P}$, where $N_P \approx 3.84$, which means a factor of six (6) faster rock disintegration on Mercury than on the Moon. The long day on Mercury is not favorable for thermal fatigue, but this is overcompensated for by higher thermal stresses.

The scaling calculation in the framework of the Delbo at al. (2014) model pertains to rocks small or comparable to $L$ only. If thermal fatigue were the only rock disintegration process, boulders larger than $L$ would survive forever: they would develop a debris talus and a protective debris layer on the top, which would isolate the inner intact part of the boulder from diurnal temperature extremes, as schematically shown in Figure 5. Thus, even if thermal failure and thermal fatigue are extremely efficient on Mercury, they cannot destroy large boulders. Other mechanisms (impacts) are needed to remove debris and expose inner parts of degrading boulder; without such mechanisms, boulder obliteration is impossible.

In summary, thermal stresses on Mercury are higher than on the Moon, and thermal fatigue is faster. These factors may cause formation of meter-scale cracks and microcracks in the first decimeters below the boulder surface, and thus aid in rock disintegration. However, these factors alone cannot explain the observed paucity of boulders on Mercury.

## 5. Conclusions

### 5.1. Summary of findings

We found that large (~5 m and larger) boulders on Mercury (more accurately, at its moderately high northern latitudes) are significantly less abundant than on the lunar highlands. We obtained a ratio of spatial density of Moon/Mercury boulder clusters to be about 30. This quantitative estimate is inherently inaccurate due to the limitation of the source data. However, our analysis firmly establishes the significant rarity of boulders on Mercury, compared to the Moon.

Higher thermal stresses and more rapid thermal fatigue on Mercury may cause rapid disintegration of the upper decimeters of the boulder surface, and thus contribute to faster boulder obliteration. However, these factors alone cannot account for the observed difference in comparison to the Moon. A thicker regolith on Mercury is likely to disproportionally reduce boulder production rate. A higher micrometeoritic flux is likely to cause micrometeoritic abrasion to be an important factor, or even to be the dominant contributor to boulder degradation, in contrast to the Moon, where degradation is dominated by larger meteoritic impacts, and disproportionally shortens the boulder life time. We believe that the latter two factors (thicker regolith and higher micrometeorite flux) acting together are responsible for the observed paucity of boulders on Mercury in comparison to the Moon.

## 5.2. Prospects

Our analysis of boulder occurrence on Mercury is limited by low quality and very limited coverage of available high-resolution images. The forthcoming BepiColombo mission to Mercury (Benkhoff et al., 2010) will provide a much wider coverage with its high resolution SIMBIO-SYS/HRIC camera (Cremonese et al., 2020). Its nominal sampling at low latitudes is ~6 m/pix, coarser than the sampling of high-resolution MDIS NAC images that we used. However, better image quality (the absence of smear, a higher signal-to-noise ratio) will favor boulder detectability; we expect that boulders detected in our survey, if they were at low latitudes, would be detectable in HRIC images. Larger images would also provide good context information for boulder occurrence, and extensive coverage will give much information. In addition, thermal measurements from the MERTIS (Mercury Radiometer and Thermal Infrared Spectrometer) instrument (Hiesinger et al., 2010, 2020) will provide estimates of the abundance of small rocks and assessment of its variability and geological associations. Thus, the results of the nominal BepiColombo mission would be very beneficial for further boulder studies. In particular, if thermal stresses contribute significantly to boulder obliteration, an anticorrelation of boulder abundance with the "hot poles" of Mercury would be expected, and this could be tested with HRIC images. Of course, observations from a lower orbit, beyond the nominal mission, with even higher resolution would be extremely useful for boulder studies.

In our discussion we saw that more robust theoretical results are needed for understanding of processes controlling boulder degradation under conditions on the surface of Mercury. A better understanding of the micrometeorite environment on Mercury can likely be achieved with the currently available observational data and modeling capacities. A more detailed theoretical treatment of thermal fatigue processes would also be beneficial.

**Acknowledgements.** We are grateful to P. Senthil Kumar and an anonymous reviewer for their positive and helpful reviews. We appreciate useful discussions with Larissa Starukhina. This paper is partly based upon work supported by NASA under award number 80NSSC17K0217 to MK. JWH acknowledges support from the NASA Lunar Reconnaissance Orbiter (LRO) Mission, Lunar Orbiter Laser Altimeter (LOLA) Experiment Team, for the study of lunar slopes and bedrock exposures, under NASA Grant 80NSSC19K0605 from the National Aeronautics and Space Administration-Goddard. MG acknowledges the Academy of Finland project no. 325806 (PlanetS) and the Russian Foundation for Basic Research project no. 19-05-00028.

**Data availability.** Source MDIS NAC images are available from the NASA Planetary Data System at https://pds-imaging.jpl.nasa.gov/. The list of MDIS NAC images with boulders is presented Table 2 in the paper.

Table 1. Summary of the survey results

|  | Images / samples surveyed | Images / samples with boulders | Percentage |
|---|---|---|---|
| The Moon | 379 | 54 | 14.3% |
| Mercury | 2993 | 14 | 0.5% |

Table 2. MDIS NAC images with boulders

| Image ID | North latitude, deg | East longitude, deg | Sampling, m/pix | Incidence angle, deg | Smear, pix |
|---|---|---|---|---|---|
| CN1066289443M | 59.42 | 311.76 | 1.64 | 61.5 | 4.5 |
| CN1066646892M | 54.82 | 284.48 | 1.33 | 62.3 | 5.7 |
| CN1066646986M | 63.19 | 288.19 | 1.24 | 67.6 | 6.1 |
| CN1066944712M | 46.44 | 260.87 | 2.28 | 63.4 | 3.3 |
| CN1067123656M | 63.96 | 254.82 | 1.02 | 73.2 | 7.5 |
| CN1067123658M | 64.13 | 254.93 | 1.03 | 73.3 | 7.4 |
| CN1067123664M | 64.65 | 255.28 | 1.08 | 73.5 | 7.0 |
| CN1067123666M | 64.83 | 255.40 | 1.10 | 73.6 | 6.9 |
| CN1067123668M | 65.00 | 255.52 | 1.11 | 73.6 | 6.8 |
| CN1067302398M | 63.74 | 241.93 | 0.94 | 75.3 | 8.1 |
| CN1067361773M | 45.52 | 230.73 | 2.27 | 72.1 | 3.3 |
| CN1067421617M | 68.63 | 237.19 | 1.50 | 78.0 | 5.0 |
| CN1067510915M | 62.28 | 226.17 | 0.82 | 77.8 | 9.4 |
| CN1067540568M | 50.11 | 219.12 | 1.39 | 76.6 | 5.4 |

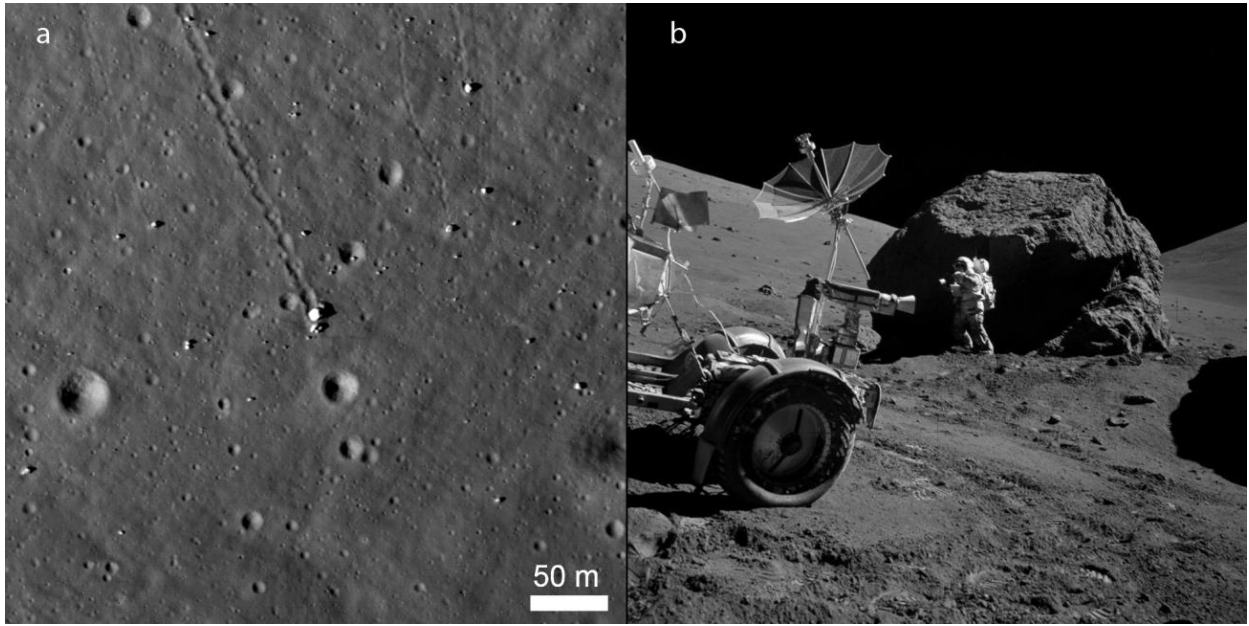

**Figure 1.** Boulders on the Moon. (a) Lunar boulders in the vicinity of Apollo-17 landing site. They rolled down from North Massif and left tracks. Portion of LROC NAC image M134991788R, illumination is from the left. (b) The biggest boulder in the center of (a), view form the west, Jack Schmitt for scale. Apollo-17 image AS17-146-22294.

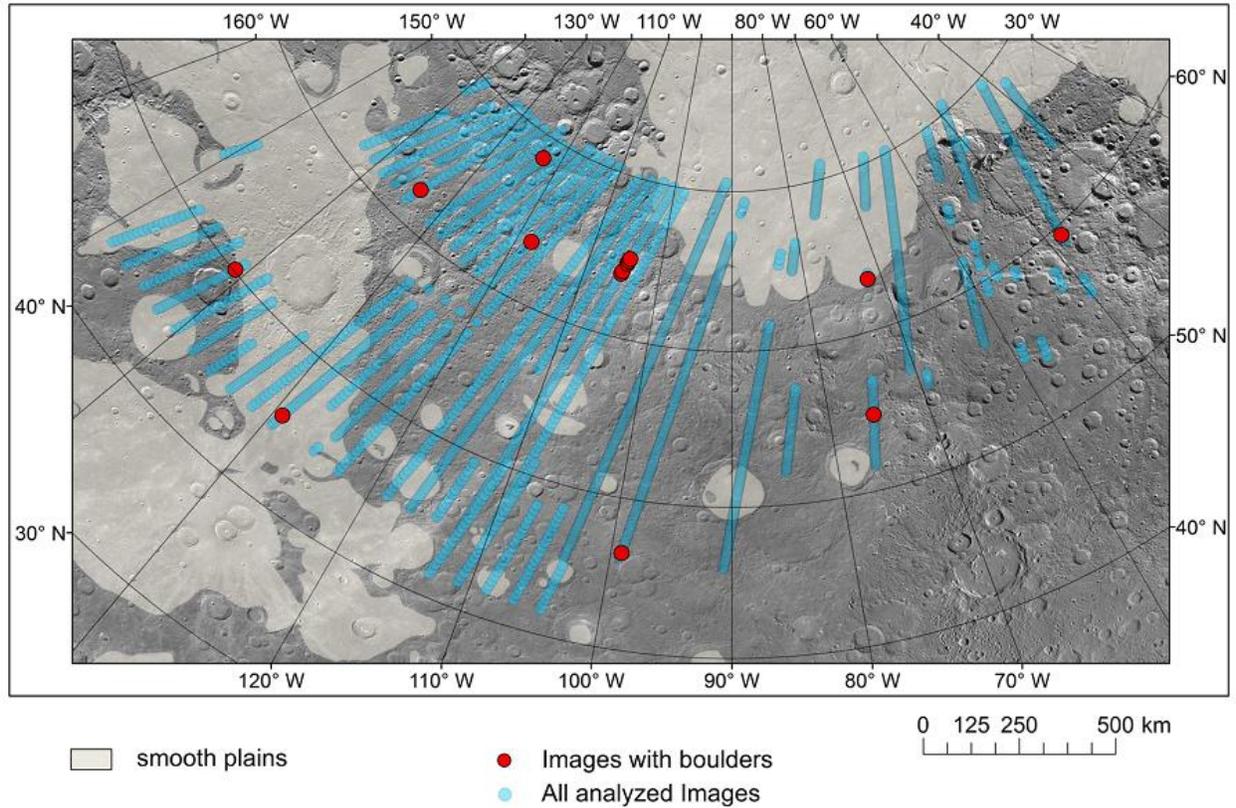

**Figure 2**. Map of a part of Mercury (Lambert azimuthal equal-area projection) showing locations of high-resolution MDIS NAC images surveyed. Small blue circles, all images surveyed; larger red circles, images containing one or more boulders. Background is low-resolution MDIS image mosaic; greyed area shows smooth plains according to Denevi et al. (2013).

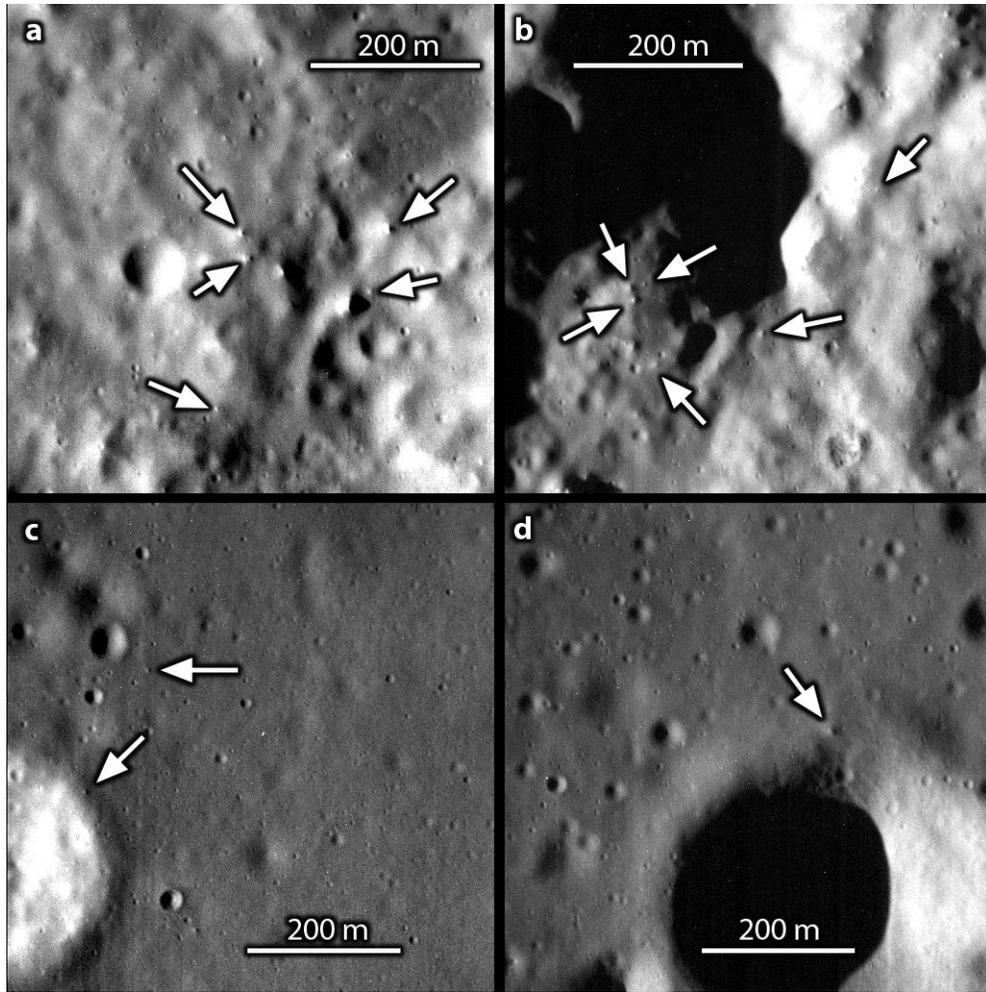

**Figure 3.** Boulders on Mercury: **(a, b)** boulders on the floor of a young large (~35 km) crater; arrows show examples of boulders; MDIS NAC images CN1067123658M and CN1067123664M; **(c)** two boulders (arrows) associated with a small (~0.3 km) crater, MDIS NAC image CN1066646986M; **(d)** single boulder (arrow) near the rim of a small (~0.4 km) crater, MDIS NAC image CN1067302398M.

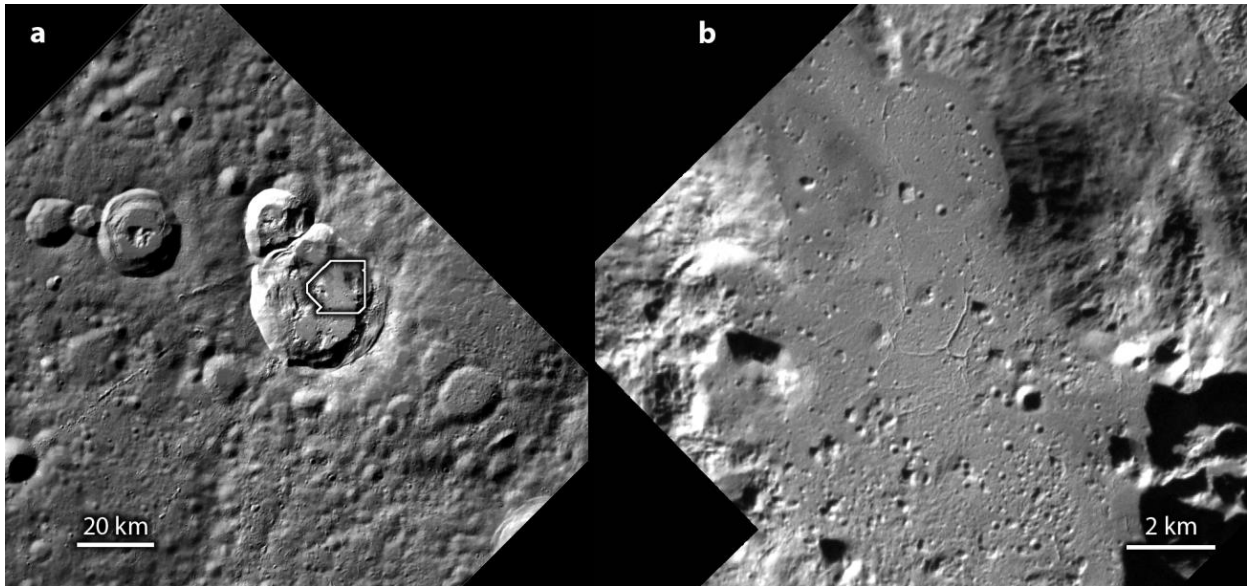

**Figure 4**. Unnamed ~35-km young crater on Mercury at 64.5°N 104.6°W. **(a)** Portion of MDIS WAC image CW0211894822G. Local equirectangular projection. **(b)** Portion of MDIS NAC mosaic N01_005261_1615539 for the area outlined in (a) showing crisp cracks in the impact melt on the crater floor. Local orthographic projection.

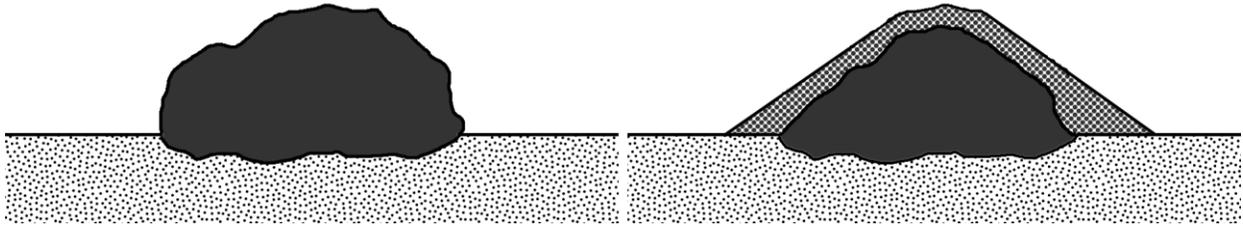

**Figure 5**. Scheme showing the ultimate effect of thermal fatigue on a large boulder. The near-surface layer of the original boulder (left) is disintegrated into debris, the debris falls down from the steeply tilting parts of the boulder surface, newly exposed parts of the intact boulder material are disintegrated into debris, and the process continues until a talus cone is formed and the remaining intact part of the boulder is completely buried under a debris layer (right). At this point development of thermal fatigue stops, and some other mechanisms are needed for further boulder degradation.